\documentclass[aps,pra,reprint,showpacs,superscriptaddress]{revtex4-1} 
\usepackage{graphicx}
\usepackage{amsmath,mathtools,verbatim}
\usepackage{amssymb}
\usepackage{mathtools}
\usepackage{amsmath}
\usepackage{bm} 
\usepackage{epstopdf}
\usepackage{braket}
\usepackage[colorlinks=true, citecolor=blue, urlcolor=blue ]{hyperref}
\usepackage{xcolor}
\usepackage{hyperref}
\usepackage{xspace} 
\usepackage{import}

\newcommand{\hide}[1]{}

\newcommand{\be}{\begin{equation}}
\newcommand{\ee}{\end{equation}}
\newcommand{\bqa}{\begin{eqnarray}}
\newcommand{\eqa}{\end{eqnarray}}

\begin{document}

\title{Photon-photon correlation of condensed light in a microcavity}
%
\author{Yijun Tang}\affiliation{Physics Department, Blackett Laboratory, Imperial College London, Prince Consort Road, SW7 2AZ, United Kingdom}
\author{Himadri S. Dhar}\affiliation{Department of Physics, Indian Institute of Technology Bombay, Mumbai 400076, India}
\author{Rupert F. Oulton}\affiliation{Physics Department, Blackett Laboratory, Imperial College London, Prince Consort Road, SW7 2AZ, United Kingdom}
\author{Robert A. Nyman}\affiliation{Physics Department, Blackett Laboratory, Imperial College London, Prince Consort Road, SW7 2AZ, United Kingdom}
\author{Florian Mintert}\affiliation{Physics Department, Blackett Laboratory, Imperial College London, Prince Consort Road, SW7 2AZ, United Kingdom}

\date{\today}    

\begin{abstract}
{The study of temporal coherence in a Bose-Einstein condensate of photons can be challenging, especially in the presence of correlations between the photonic modes.
In this work, we use a microscopic, multimode model of photonic condensation inside a dye-filled microcavity and the quantum regression theorem, to derive an analytical expression for the equation of motion of the photon-photon correlation function.
{This allows us to derive the coherence time of the photonic modes and identify a nonmonotonic dependence of the 
temporal coherence of the condensed light with the cutoff frequency of the microcavity.}}
%
\end{abstract}     
\maketitle

\section{Introduction\label{intro}}

A defining property of a Bose-Einstein condensate (BEC) is the macroscopic coherence
exhibited by the particles in the lowest energy mode -- a feature that allows the condensate to behave like a massive quantum wave~\cite{Ketterle2002,Pitaevskii2003}. While features such as thermal equilibrium and large population of the ground state are the tell-tale signatures of a BEC, onset of quantum coherence can be a defining characteristic of condensates that do not thermalize completely and essentially operate out of equilibrium, such as exciton-polaritons~\cite{Kasprzak2006, Balili2007} and photons~\cite{Klaers2010,Marelic2015,Greveling2018,Schofield2023}. As such, theoretical and experimental investigation of coherence in both equilibrium and non-equilibrium condensates plays an important role in characterizing the properties of these macroscopic states.  
Over the years, coherence of BEC has been observed using interference experiments with ultracold atoms~\cite{Andrews1997,Bloch2000}. Moreover, coherence in condensates of excitons-polaritons~\cite{Deng2007} and organic polaritons~\cite{Plumhof2014} have also been reported, while spontaneous phase selection~\cite{Schmitt2016} and spatio-temporal coherence~\cite{Marelic2016,Damm2017} has been observed in photonic condensates. 

The photonic Bose-Einstein condensates formed inside a dye-filled microcavity 
is driven-dissipative in nature, sustained by a detailed balance between the rate at which the dye molecules are incoherently driven and losses of molecular and photonic excitation in the system. 
{The thermalization  of the photon gas inside the microcavity is due to the vibrational states of the dye molecule, which allow thermal equilibration of photons via energy-dependent absorption and emission processes.}
%
From a physical perspective, one of the central differences between a photonic BEC and a laser is the rate of thermalization it is operating under.
While a photonic BEC works in a near-equilibrium regime, close to thermal equilibrium with the dye molecules,
a laser operates at a much lower thermalization rate and is firmly in the non-equilibrium regime. As such, macroscopic occupation of photons occurs in the lowest energy mode in a photon condensate, irrespective of the gain, whereas for a laser the large population corresponds to the mode with the highest gain. In between these regimes, lie exotic phases of light that exhibit strong multimode properties~\cite{Walker2018}, non-stationary kinetics~\cite{Schmitt2015,Walker2020} and possible vortex-like features~\cite{Dhar2021}.

A key characteristic of Bose-Einstein condensates that operate in the quasiequilibrium regime is the spatio-temporal coherence. While condensates of exciton-polaritons exhibit significant inter-particle interactions, which lead to phase coherence~\cite{Roumpos2012} and superfluid behavior in the system~\cite{Carusotto2013}, photons in a BEC do not interact strongly with each other~\cite{Nyman2014, Alaeian2017}. Instead coherence is created from indirect interaction of photons with the dye molecules, mediated via stimulated emissions~\cite{Snoke2013}.
This leads to 
coherence properties such as symmetry-broken phase coherence~\cite{Schmitt2016}, grand-canonical photon statistics~\cite{Schmitt2014} and transition from short-range to long-range spatial order across the condensation threshold~\cite{Marelic2016,Damm2017} that have been experimentally observed.
While phase coherence is
generally well understood in these systems, studies on temporal coherence which is consistent with experimental results have been limited~\cite{Marelic2016,Walker2018}. 

The properties of a photon condensation is well described by a microscopic model~\cite{Kirton2013}, derived from the light-matter interaction between a multimode cavity and an ensemble of emitters that represent the electronic and vibrational states of the dye molecules. The model can perfectly capture the thermalization~\cite{Kirton2015} and quasi-equilibrium properties~\cite{Keeling2016} of the photon gas in the cavity, and also highlight 
features such as  decondensation~\cite{Hesten2018} and noncritical slowing down of dynamics~\cite{Walker2019}.
First order correlation function and the photon linewidth~\cite{Kirton2015} based on a single-mode model predict a perfect temporal coherence in the condensed mode, consistent with the Schawlow-Townes limit at large photon numbers~\cite{Schawlow1958}.
While the theoretical results are in agreement with experimental observations in the vicinity of the BEC threshold~\cite{Marelic2016,Walker2018}, it exhibits significant deviation at higher photon numbers. 

In this work, we derive the equation of motion of the photon-photon or the first order correlation function of the photon gas inside a dye-filled microcavity, especially
when interactions between the different cavity modes
are explicitly taken into account.
Such a model already provides great insight into the role of spatial coherence in the kinetics of the condensed light in the cavity~\cite{Keeling2016,Dhar2021}. Our focus here is on the temporal coherence of the condensed light, and by using the quantum regression theorem, we derive an analytical expression for the equation of motion of the photon-photon correlation. This allows us to study the temporal coherence or the coherence time of light inside the cavity for different system parameters.


The paper is arranged as follows. After the Introduction in Sec.~\ref{intro}, we study the multimode model in Sec.~\ref{model}, and derive the rate equations for the photon correlations and molecular excitations. In Sec.~\ref{sec:corr_func}, we represent the photon-photon correlations in terms of the coefficients of the density matrix of the system.
In Sec.~\ref{coefficients}, the time derivative of these coefficients are 
obtained to ultimately
derive the equation of motion of the photon-photon correlation~\ref{sec:eom}. In Sec.~\ref{temporal}, the time evolution of the correlation is numerically studied and the coherence time of the condensed light for different cutoff frequencies and pumping rate are presented. The results are discussed in Sec.~\ref{conclusion}.

%


\section{Theoretical Model\label{model}}

The interaction of photons with the dye molecules inside a microcavity can be studied using a microscopic model~\cite{Kirton2013}. In its most general form, the theoretical model is valid for multiple cavity modes and also takes into account finite intermode correlations~\cite{Keeling2016}. The dynamics of the quantum system is governed by the following Master equation,
\begin{eqnarray}
\frac{d\rho}{dt} &=& -i[\hat{H_{0}},\hat{\rho}]+ \frac{1}{2}\sum_{i,p} (\kappa L[\hat{a}_p]+\Gamma_{\uparrow}^i L
[\hat{\sigma}^{+}_{i}]+\Gamma_\downarrow L[\hat{\sigma}^{-}_{i}]) \nonumber\\
&+&\frac{1}{2}\sum_{i,p,q} (\Psi_{p,q}^i \{\mathcal{A}_{q}[\hat{a}_q\hat{\sigma}^{+}_{i}\rho,\hat{a}^{\dagger}_{p}\hat{\sigma}^{-}_{i}]
+ \mathcal{E}_{p}[\hat{a}^{\dagger}_p\hat{\sigma}^{-}_{i}\rho,\hat{a}_q\hat{\sigma}^{+}_{i}]\} \nonumber\\ \nonumber\\
&+& \textrm{h.c.}),
\label{master}
\end{eqnarray}
where $\hat{H}_0 = \sum_{p} \delta_p\hat{a}_p^\dag \hat{a}_p$ is the bare energy of the cavity photons, and $L[\hat{x}]=2\hat{x}\rho\hat{x}^\dag-\{\hat{x}^\dag\hat{x},\rho\}$, with $\{\cdot\}$ being the anticommutator. 
Here, $\hat{a}_p$ ($\hat{a}^\dag_p$) is the annihilation (creation) operator of photons in the $p^\textrm{th}$ cavity mode, and $\kappa$ is the rate at which it is lost from the cavity.
The Pauli operators $\sigma^{\pm}_i$ denote the electronic states of the dye molecule at location $\textbf{r}_i$ in the cavity plane,  
which is pumped at a rate $\Gamma_\uparrow^i$, but decays with a uniform rate $\Gamma_\downarrow$ to non-cavity modes. The rate of absorption and emission of photons in the $p^\textrm{th}$ mode is given by $\mathcal{A}_{p}$ and $\mathcal{E}_{p}$, respectively.
The expression   $\Psi_{p,q}^i = \psi_{p}(\bold{r}_{i})\psi_{q}(\bold{r}_{i})$
is given in terms of the mode functions $\psi_{p}(\bold{r}_i)$ of cavity with quantum number $p$.
The absorption and emission rates for the dye molecules, $\mathcal{A}_{p}$ and $\mathcal{E}_{p}$, can either be calculated~\cite{Kirton2015} or estimated from experimental data of the absorption crosssection~\cite{Nyman2017}.
Moreover, these rates are known to follow the Kennard-Stepanov relation~\cite{Kennard1918,Kennard1926,Stepanov1957}, $\mathcal{A}_{p}$ = $\mathcal{E}_{p} e^{-\beta\delta_p}$, where $\delta_p = \omega_\textrm{ZPL} - \omega_p$, with $\omega_p$ being the frequency of mode $p$ and $\omega_\textrm{ZPL}$ is the zero-phonon line.


The dynamics as well as the steady behavior of the cavity modes and the molecular excitation can be studied
in terms of the equations of motion of the relevant observables such as the mode population or photon correlation from the above master equation. In most cases, it is easier to work with rate equations compared to directly solving the density matrix, especially when the system contains large number of modes and an inhomogeneous distribution of molecular excitations.

On the other hand, equations of motion for observables such as the mode population $\langle \hat{a}^\dag_p\hat{a}_p\rangle$ and the molecular excitation $\langle\sigma^+_i\sigma^-_i\rangle$, or two-mode correlation function  $\langle \hat{a}^\dag_p\hat{a}_q\rangle$  can be computed with relatively 
little computational resources when certain approximations are taken into consideration.
First, 
it is helpful to use the fact that coupling between different dye-molecules have been neglected on account of the incessant collision between the molecules of the dye and the solvent, which quickly decoheres any intermolecule coherence. Secondly, one can invoke the semiclassical approximation such that correlations between molecules and photons can be factorized i.e., $\langle\sigma^+_i(t)\hat{a}_q(t)\rangle \approx \langle\sigma^+_i(t)\rangle\langle\hat{a}_q(t)\rangle$. This is typically a good approximation when the number of emitters in a cavity is large, which is true in the case of dye-filled microcavity, where the number of dye molecules inside the microcavity is very large. 

The photon population and correlations can be written as a matrix $\mathbf{n}$, with elements $n_{pq}(t)=\langle \hat{a}^\dag_p(t)\hat{a}_q(t)\rangle$, and the molecular excitation fraction as a vector $\textbf{f}$ with elements $f_i=\sum_j \langle\sigma^+_j\sigma^-_j\rangle\delta(\mathbf{r}_i-\mathbf{r}_j)$. 
The semiclassical equations of motion for $\mathbf{n}$ and $\mathbf{f}$  are then given by, 
\begin{eqnarray}
\dot{\textbf{n}} &=& \left( i\mathbf{\Omega}-\frac{\kappa}{2}\right) \textbf{n} +  \{\textbf{f}^{+}~\textbf{E}(\textbf{n}+\mathbb{I}) - \textbf{f}^{-}\textbf{A}\textbf{n}\}+ \mathrm{h.c.},\label{eq:ndot}~ \\
\dot{\textbf{f}} &=&  - \{\Gamma_\downarrow + 2 \tilde{\mathbf{E}}\} \textbf{f}
+\{\mathbf{\Gamma}_\uparrow + 2 \tilde{\mathbf{A}}\} (\mathbf{1} - \textbf{f}),\label{eq:fdot}
\end{eqnarray}
where $\textbf{f}^{+}$ and $\textbf{f}^{-}$ have elements $f^+_{pq}=\sum_i f_i\Psi_{pq}^i$ and $f^-_{pq}=\sum_i (1-f_i)\Psi_{pq}^i$, respectively. 
$\mathbf{A}, \mathbf{E}$ and $\mathbf{\Omega}$ are diagonal matrices with elements $\mathcal{A}_p, \mathcal{E}_p$ and $\omega_p$, respectively. $\mathbf{\Gamma}_\uparrow$ is simply a vector with terms ${\Gamma}_\uparrow^i$. 
The matrices, $\tilde{\mathbf{E}}$ and $\tilde{\mathbf{A}}$ are diagonal with elements 
$\mathrm{Tr}[\mathbf{\Phi}_i\mathbf{E}(\mathbf{n} + \mathbb{I})]$ and
$\mathrm{Tr}[\mathbf{\Phi}_i\mathbf{n}\mathbf{A}]$, respectively, where $\mathbf{\Phi}_i$ has the same dimension as $\mathbf{n}$ and has elements $\Psi_{pq}^i$. Importantly, Eqs.~(\ref{eq:ndot}) and (\ref{eq:fdot}) allows for the computation of the photon correlation function $\langle \hat{a}^\dag_p(t)\hat{a}_q(t)\rangle$ at time $t$, as well as at the steady state of the system, for a given set of system parameters.

\section{Photon-photon correlation function\label{sec:corr_func}}

The primary focus of the work is to calculate the first-order correlation function, 
which would allow for the estimation of the spectral linewidth, as well as the temporal coherence of the multimode photon gas inside the dye-filled microcavity. The key quantity of interest is the photon-photon correlation,  
\begin{equation}
c_{pq}(t_2-t_1) = \langle\hat{a}_p^\dagger(t_2)\hat{a}_{q}(t_1)\rangle,
\end{equation}
where $p$ and $q$ denotes the cavity modes. For most experiments involving the investigation of spatio-temporal coherence, the initial state of the system at $t_1$ is
the steady state $\rho_{ss}$.
 As such only the time difference, $t = t_2-t_1$ is relevant and the two-time correlation reduces to $c_{pq}(t) = \langle\hat{a}_p^\dagger(t)\hat{a}_{q}(0)\rangle$, where one can set the initial time $t_1=0$, and $t_2=t$. Therefore, the two-time correlation can now be calculated using the term $\text{Tr}[\hat{a}_p^\dagger(t)\hat{a}_{q}(0)\rho_{ss}]$.

The correlation function can be calculated using the quantum regression theorem~\cite{Carmichael1993,Gardiner2000}, which allows for setting up an equation of motion for the two-time correlation. The function  
$\langle\hat{a}_p^\dagger(t)\hat{a}_{q}(0)\rangle$ can considered as the time-evolution of the term  $\langle\hat{a}_p^\dagger\rangle$, governed by the master equation given in Eq.~(\ref{master}), however, with the initial state given by $\rho'(0)=\hat{a}_q \rho_{ss}$. In other words, using the regression theorem, we have the relation,
\begin{eqnarray}
c_{pq}(t) = \langle\hat{a}_p^\dagger\rangle = \textrm{Tr}[\hat{a}_p^\dagger\rho']= \textrm{Tr}[\hat{a}_p^\dagger(t)\hat{a}_{q}(0)\rho_{ss}].
\end{eqnarray}
Similar approaches have been used to derive the first order~\cite{Kirton2015} and second order correlation function~\cite{Walker2020}. But these are mostly restricted to just single-mode systems, where there are no interactions arising from intermode correlations and the overall dynamics is fairly simple. 


To apply the regression theorem, it is helpful to work with the density matrix formalism, which can be written in an expanded form using an appropriate orthonormal basis. For $M$ photon modes and $N$ molecules inside the cavity, one such basis is $\{|\underline{n},\underline{s}\rangle\}$, where $|\underline{n}\rangle = |n_0 n_1\dots n_p \dots n_M\rangle$ with $n_p$ being the population of the $p^\text{th}$ mode. 
Similarly, $|\underline{s}\rangle=|s_1 s_2 \dots s_i\dots s_N\rangle$
is the molecular excitation state, where $s_i$ can take values $0$ or $1$, depending on whether the $i^\text{th}$ molecule is excited or not. 
Thus, the density matrix $\rho$ can then be expressed in this basis as
\begin{equation}
\rho=\sum_{\underline{n},\underline{n}',\underline{s}} C_{\underline{n},\underline{n}',\underline{s}}|\underline{n}\rangle\langle\underline{n}'|\otimes |\underline{s}\rangle\langle\underline{s}|,
\end{equation}
where $|\underline{n},\underline{s}\rangle\langle\underline{n}',\underline{s}'|=|\underline{n}\rangle\langle\underline{n}'|\otimes |\underline{s}\rangle\langle\underline{s}|$,
based on the fact that there are no coherence between the molecules and the semiclassical approximation is valid. 
%
%
Now, the diagonal elements of $\rho$ are those that satisfy $\underline{n}=\underline{n}'$ and the off-diagonal terms are for $\underline{n}\neq\underline{n}'$. In general, $\underline{n},\underline{n'}  \geq 0$, which means $n_p \geq 0~\forall~p$ and implies that mode population is non-negative. As such, the following rearrangement of terms can be made, 
\begin{eqnarray}
\sum_{\underline{n}'\geq 0} \sum_{\underline{n}\geq 0} |\underline{n}\rangle\langle\underline{n}'| &=& \sum_{\underline{n}'\geq 0} \sum_{\underline{n}'-\underline{k}\geq 0} |\underline{n}'-\underline{k}\rangle\langle\underline{n}'| \nonumber\\
&\equiv& \sum_{\underline{k} \leq \underline{n}}\sum_{\underline{n}\geq 0}  |\underline{n}-\underline{k}\rangle\langle\underline{n}|
\end{eqnarray}
So, $\rho$ can now be rearranged in terms of $|\underline{n}-\underline{k}\rangle\langle n|$, where $\underline{n'}-\underline{n}=\underline{k}$, 
\begin{equation}
\rho=\sum_{\underline{k}}\bigg(\sum_{\underline{n},\underline{s}}R^{\underline{k}}_{\underline{n},\underline{s}}|\underline{n}-\underline{k}\rangle\langle\underline{n}|\otimes |\underline{s}\rangle\langle\underline{s}|\bigg),
\label{expandfull}
\end{equation}
and $R^{\underline{k}}_{\underline{n},\underline{s}}=\bra{\underline{n}-\underline{k},\underline{s}}\rho\ket{\underline{n},\underline{s}}$.
Note that $\underline{k}$ can be non-positive and the above expression 
can be thought of as representing the density matrix by summing the $k^{\textrm{th}}$ diagonal terms. 

The two-time correlation can now be computed by taking the trace 
$\text{Tr}[\hat{a}_{p}^\dag\rho']$, where  
\begin{equation}
\rho' = \hat{a}_q\rho= \sum_{\underline{k},\underline{n},\underline{s}}P^{\underline{k}}_{\underline{n},\underline{s}}~|\underline{n}-\underline{k}\rangle\langle\underline{n}|\otimes |\underline{s}\rangle\langle\underline{s}|, \label{rhoprime}
\end{equation}
where we expand $\rho'$ similarly as in Eq.~(\ref{expandfull}) with new set of coefficients $\{P^{\underline{k}}_{\underline{n},\underline{s}}\}$ which are related to the steady state coefficients by $ P^{\underline{k}}_{\underline{n},\underline{s}}(0)=\sqrt{n_q-k_q+1}~ R^{\underline{k}-\underline{k}_q}_{\underline{n},\underline{s}}$, where $\underline{k}_{q}$ is defined as a vector with the $q^\text{th}$ term as unity and zero elsewhere. 
As such, we obtain the following expression,
%

\begin{eqnarray}
c_{pq}(t) &=& \text{Tr}[\hat{a}_{p}^\dag\rho'] = \text{Tr}\bigg[\hat{a}_{p}^\dag
\sum_{\underline{k},\underline{n},\underline{s}}~P^{\underline{k}}_{\underline{n},\underline{s}}|\underline{n}-\underline{k}\rangle\langle\underline{n}|\otimes |\underline{s}\rangle\langle\underline{s}|\bigg] \nonumber\\
&=& \sum_{\underline{n},\underline{s}}
\bigg( \sum_{\underline{k},\underline{n}'}
\sqrt{n_p-k_p+1} ~P^{\underline{k}}_{\underline{n},\underline{s}}\langle \underline{n}'|\underline{n}-\underline{k}+\underline{k_p}\rangle \nonumber\\
&\times& 
\langle\underline{n}|\underline{n}'\rangle \bigg)
\otimes \sum_{\underline{s}'} \langle \underline{s}'|\underline{s}\rangle\langle\underline{s}|\underline{s}'\rangle \nonumber\\
&=&
\sum_{\underline{n},\underline{s}}\sqrt{n_p} ~P^{\underline{k}_p}_{\underline{n},\underline{s}}. 
\label{model1}
\end{eqnarray}
%
%
This gives us a time-evolution of the photon-photon correlation in terms of the coefficients, 
\begin{equation}
\dot{c}_{pq}(t) = \frac{d}{dt}\langle\hat{a}_p^\dagger(t)\hat{a}_{q}(0)\rangle=\sum_{\underline{n},\underline{s}}\sqrt{n_{p}}~\dot{P}^{\underline{k_{p}}}_{\underline{n},\underline{s}}.\label{rfregression}
\end{equation}


\section{Calculation of coefficients\label{coefficients}}
The estimation of the photon-photon correlation now depends on finding the solution to Eq.~(\ref{rfregression}), which is obtained by finding the terms  $\dot{P}^{\underline{k}_p}_{\underline{n},\underline{s}}$.
In particular, the above time derivative needs to be expressed as a function of the coefficients 
defined in Eqs.~(\ref{expandfull}) and (\ref{rhoprime}), which 
gives the expression for the operator $\rho'$ in terms of $\{P^{\underline{k}_p}_{\underline{n},\underline{s}}\}$.
Hence, to obtain the time-derivative of the two-time correlation, the natural step 
is to calculate the derivative of the operator $\rho'$ using the master equation in Eq.~(\ref{master}). In particular, the different terms in the master equation needs to be investigated for their contribution to the time-derivative, starting from the Hamiltonian $\hat{H}_0$ to the Lindblad terms $L[\hat{\sigma}^\pm_i]$ and $L[\hat{a}_p]$. For example, contribution from $\hat{H}_{0}$ in finding the relevant terms in $\dot{\rho}$ will simply come from 
\begin{equation}
\dot{\rho}\overset{\hat{H}_0}{=}-i[\hat{H_{0}},\rho] = -i(\hat{H}_{0}\rho-\rho\hat{H}_{0}),
\end{equation}
where $\hat{H}_0 = \sum_{m} \hat{a}_m^\dag \hat{a}_m$. Here are the contributions from the different terms in the master equation:\\

\noindent $\bullet$ $\hat{H}_{0}$ contribution
\begin{equation}
\dot{P}^{\underline{k}_p}_{\underline{n},\underline{s}}\overset{\hat{H}_0}{\longrightarrow}i\delta_{p}P^{\underline{k}_p}_{\underline{n},\underline{s}}.\\
\label{H}
\end{equation}

\noindent $\bullet$ $L[\hat{a}_p]$ contribution
\begin{equation}
\dot{P}^{\underline{k}_p}_{\underline{n},\underline{s}}\overset{L[\hat{a}_p]}{\longrightarrow}\sum_{m}\kappa\bigg[C_{0}(m)P^{\underline{k}_p}_{\underline{n}+\underline{k}_{m},\underline{s}}-(n_{m}-\frac{\delta_{p,m}}{2})P^{\underline{k}_p}_{\underline{n},\underline{s}}\bigg],\label{Kappa}
\end{equation}
where $C_{0}(m)=\sqrt{(n_{m}+1)(n_{m}+1-\delta_{p,m})}$. 
Note that $\kappa$ describes the loss rate for all cavity modes, and a state $\underline{n}$ can lose a photon in mode $m$ and thus change into  $\underline{n}-\underline{k}_{m}$, where $\underline{k}_{m}$ (as defined earlier) is a vector with the $m^\text{th}$ element as 1 and 0 everywhere else. Conversely a photon population $\underline{n}$ can come from state $\underline{n}+\underline{k}_{m}$ by losing a photon in mode $m$. 
As such, these states have probability $P^{\underline{k}_p}_{\underline{n}+\underline{k}_{m},\underline{s}}$.\\

\noindent$\bullet$ $L[\hat{\sigma}^{\pm}_{i}]$ contribution
\begin{equation}
\dot{P}^{\underline{k}_p}_{\underline{n},\underline{s}}\overset{L[\hat{\sigma}^{+}_{i}]}{\longrightarrow}
\sum_{i,\forall s_i = 1} \Gamma_{\uparrow}^i~P^{\underline{k}_p}_{\underline{n},\underline{s}-\underline{s}_i}-\tilde{\Gamma}_{\uparrow}~
P^{\underline{k}_p}_{\underline{n},\underline{s}},
\label{P9}
\end{equation}
where $\tilde{\Gamma}_{\uparrow}=\sum_{i,\forall s_i = 0}\Gamma_{\uparrow}^i$ and 
$\underline{s}_{i}$ is a vector with the $i^\text{th}$ element as 1 and 0 everywhere else. 
As $\Gamma_{\uparrow}^i$ is the pumping rate it excites the state $\underline{s}$ to $\underline{s}+\underline{s}_{i}$ by exciting the $i^\text{th}$ molecule.
Similarly, 
a state $\underline{s}$ may be pumped from a state $\underline{s}-\underline{s}_{i}$ with probability $P^{\underline{k}_p}_{\underline{n},\underline{s}-\underline{s}_{i}}$. 
The term $L[\hat{\sigma}^{-}_{i}]$ being decay of molecule with rate $\Gamma_{\uparrow}$, simply produces a the reverse state transformation. 

A more detailed description of the above transformations is shown in Appendix~\ref{app1}. \\


\noindent$\bullet$ $\mathcal{A}_{m}$ and $\mathcal{E}_{m'}$ contribution 
\begin{align}
\dot{P}^{\underline{k}_p}_{\underline{n},\underline{s}}&\overset{\mathcal{A}_{m}}{\longrightarrow}\sum_{m, m'}\frac{\mathcal{A}_{m'}}{2}\bigg[C_{1}\bigg(\sum_{i,\forall~s_i=1}\Psi_{m,m'}^{i}~P^{\underline{k}_p-\underline{k}_{m'}+\underline{k}_{m}}_{\underline{n}+\underline{k}_{m},\underline{s}-\underline{s}_{i}}\bigg)
\nonumber\\
&- C_{2}\bigg(\sum_{i,\forall~s_i=0}\Psi_{m,m'}^{i}\bigg)P^{\underline{k}_p-\underline{k}_{m'}+\underline{k}_{m}}_{\underline{n},\underline{s}}\bigg],
\label{Pa16}
\end{align}
where $C_{1}=\sqrt{(n_{m'}-\delta_{p,m'}+1)(n_{m}+1)}$ and $C_{2}=\sqrt{(n_{m}-\delta_{m,p}+1-\delta_{m,m'})(n_{m'}-\delta_{p,m'})}$. 
%

Note that the process of absorption and emission, as governed by $\mathcal{A}_{m}$ and $\mathcal{E}_{m'}$ in Eq.~(\ref{master}), gives rise to inter-mode correlations. In the case of $m=m'$ the mode is coupled to itself. Mathematically, this is a state with probability $P^{\underline{k}_p}_{\underline{n},\underline{s}}$, which upon absorption of a photon in mode $m$ by the $i^\text{th}$ molecule at location $\textbf{r}_{i}$, will transform to a state with probability $P^{\underline{k}_p}_{\underline{n}+\underline{k}_{m},\underline{s}-\underline{s}_{i}}$.
Now, including $m\neq m'$, introduces inter-mode correlations, which reveals itself in the coupling of different $\underline{k}_i$ terms. For instance, $\underline{k}_p$ in the derivative on the left hand side is connected to the $\underline{k}_p-\underline{k}_{m'}+\underline{k}_{m}$ on the right for all $m$ and $m'$.
%
Similar calculations can also be done for transformations arising from the emission term $\mathcal{E}_{m'}$. The contribution from Hermitian conjugates in Eq.\ref{master} are calculated in similar manner. 

Now, the full expression of $\dot{P}^{\underline{k}_p}_{\underline{n},\underline{s}}$ is simply the sum of the different contributions, which gives us a general relation of the time derivative to the set $\{P^{\underline{k}_p}_{\underline{n},\underline{s}}\}$.

\section{Equation of motion\label{sec:eom}}

In Eq.~(\ref{rfregression}), we represent the time-derivative of the two-time correlation function, $\dot{c}_{pq}(t)=\frac{d}{dt}\langle\hat{a}_p^\dagger(t)\hat{a}_{q}(0)\rangle$ in terms of the rate of change of the probabilities  $\dot{P}^{\underline{k}_p}_{\underline{n},\underline{s}}$. Moreover, in Eqs.~(\ref{H})-(\ref{Pa16}), we derived the time-derivatives in terms of the probabilities $\{{P}^{\underline{k}_p}_{\underline{n},\underline{s}}\}$. 
As such, the equation of motion of the two-time correlation function $\langle\hat{a}_p^\dagger(t)\hat{a}_{q}(0)\rangle$ can now be derived independently of these probabilities, which in practical conditions can be very difficult to  estimate. 



From Sec.~\ref{coefficients}, the contributions from the $\hat{H}_{0}$ and $L[\hat{a}_p]$ terms in the master equation lead to

\begin{equation}
\frac{d}{dt}\langle\hat{a}_p^\dagger(t)\hat{a}_{q}(0)\rangle =
(i\delta_{p} -{\kappa}/{2})\langle\hat{a}_p^\dagger(t)\hat{a}_q(0)\rangle,
\end{equation}

and therefore
introduce an oscillatory and a decay term in the equation of motion. 
The terms related to pumping and decay of molecules in the system, given by $L[\sigma^{\pm}_i]$, do not contribute to the photon correlation. 
However, the absorption and emission terms, given by $\mathcal{A}_{m}$ and $\mathcal{E}_{m'}$, make a significant contribution to the equation of motion. For $\mathcal{A}_{m}$ this is given by
\begin{equation}
\frac{d}{dt}\langle\hat{a}_p^\dagger(t)\hat{a}_{q}(0)\rangle =
-\sum_{\underline{n},\underline{s},m}\frac{\mathcal{A}_{m}\sqrt{n_{m}}}{2}\bigg(\sum_{i, \forall s_i=0}\Psi^i_{p,m}\bigg)P^{\underline{k}_{m}}_{\underline{n},\underline{s}}.
\label{AM}
\end{equation}

The contribution from $\mathcal{E}_{m'}$ is given simply by replacing absorption term with emission, the index $m$ by $m'$, and the summation over all $s_{i}=1$. An extended derivation of Eq.~(\ref{AM}) is shown in Appendix~\ref{app2}.

Importantly, the contributions from $\mathcal{A}_{m}$ and $\mathcal{E}_{m'}$ still contains coefficients $P^{\underline{k}_{m}}_{\underline{n},\underline{s}}$.
A useful condition to use at this point is the semiclassical approximation discussed in Sec.~\ref{model}, which factorizes the correlation between photons and molecules. This leads to $P^{\underline{k}_{p}}_{\underline{n},\underline{s}}\approx P^{\underline{k}_p}_{\underline{n}}P_{\underline{s}}$, where one can interpret $P_{\underline{s}}$ to be probability for molecules to have an excitation profile described by $\underline{s}$. As such Eq.~(\ref{model1}) can be written as, 
\begin{eqnarray}
{c}_{pq}(t) &=& \langle\hat{a}_p^\dagger(t)\hat{a}_{q}(0)\rangle=\sum_{\underline{n},\underline{s}}\sqrt{n_{p}}~P^{\underline{k}_{p}}_{\underline{n},\underline{s}}\nonumber\\
&=& \sum_{\underline{n}}\sqrt{n_{p}}P^{\underline{k}_{p}}_{\underline{n}}\sum_{\underline{s}}P_{\underline{s}}
= \sum_{\underline{n}}\sqrt{n_{p}}P^{\underline{k}_{p}}_{\underline{n}}
%
\label{semiclassical}
\end{eqnarray}
where the sum over probabilities of all possible molecular excitation profile is unity, i.e., $\sum_{\underline{s}}P_{\underline{s}}=1$. 
Now, to simplify the expression in Eq.~(\ref{AM}), we use the semiclassical approximation in Eq.~(\ref{semiclassical}), such that 
\begin{eqnarray}
\dot{c}_{pq}(t) &\approx& -\sum_{\underline{n},m}\frac{\mathcal{A}_{m}\sqrt{n_{m}}}{2}~P^{\underline{k}_{m}}_{\underline{n}}
\sum_{\underline{s}}\bigg(P_{\underline{s}}
\sum_{i,\forall s_i=0}\Psi^i_{p,m}\bigg),\nonumber\\
&=& -\sum_{m}\frac{\mathcal{A}_{m}}{2}~c_{mq}(t)
\sum_{\underline{s}}\bigg(P_{\underline{s}}\sum_{i,\forall s_i=0}\Psi^i_{p,m}\bigg).
\end{eqnarray}
%
The coefficient $\sum_{\underline{s}}(P_{\underline{s}}\sum_{i,\forall s_i=0}\Psi^i_{p,m})$, here is not straightforward. It is a sum of the mode function $\Psi^i_{p,m}$ at locations $\mathbf{r}_i$, where the molecule is not excited as described by the vector $|\underline{s}\rangle$. The above expression can be re-written as 
shown in the Appendix~\ref{app3}, as $\sum_i \Psi^i_{p,m}(1-f_i)$, where $f_i$ is the probability of finding an excited molecule at location $\mathbf{r}_i$ or the excitation fraction. A similar expression can also be obtained from the emission $\mathcal{E}_{m'}$ term. 


Therefore, bringing all the terms together, the full equation of motion of for the photon-photon correlation is given by 
%
\begin{align}
\frac{d}{dt}\langle\hat{a}_p^\dagger(t)&\hat{a}_{q}(0)\rangle = (i\delta_{p}-{\kappa/}{2})\langle\hat{a}_p^\dagger(t)\hat{a}_{q}(0)\rangle  \nonumber\\
&+\sum_{m}\bigg(\frac{\mathcal{E}_{m}}{2}f^{+}_{mp}-\frac{\mathcal{A}_{m}}{2}f^{-}_{mp}\bigg)\langle\hat{a}_m^\dagger(t)\hat{a}_{q}(0)\rangle, 
\label{FULL22}
\end{align}
where $f^{+}_{mp}=\sum_{i}f_{i}\Psi_{m,p}^i$ and $f^{-}_{mp}=\sum_{i}(1-f_{i})\Psi_{m,p}^i$. \\

%
%
{The photon-photon correlation or the first order correlation function allows for the estimation of the emission spectrum of the cavity and the temporal coherence of the different modes. 
The time evolution of the correlation function $c_{pq}(t)=\langle\hat{a}_p^\dagger(t)\hat{a}_{q}(0)\rangle$ is governed by the equation of motion in Eq.~(\ref{FULL22}), which can be numerically solved for a particular set of system parameters. The initial state of the system, at time $t=0$, is the steady state correlation given by $\langle\hat{a}^\dag_p(0)\hat{a}_q(0)\rangle$ = $\langle\hat{a}_p^\dag\hat{a}_q\rangle_{ss}$, which is obtained by finding the the steady state solutions of Eqs.~(\ref{eq:ndot}) and (\ref{eq:fdot}) discussed in Sec.~\ref{model}.}

\section{Temporal coherence\label{temporal}}


\begin{figure}[t] 
\center
\includegraphics[width=0.44\textwidth]{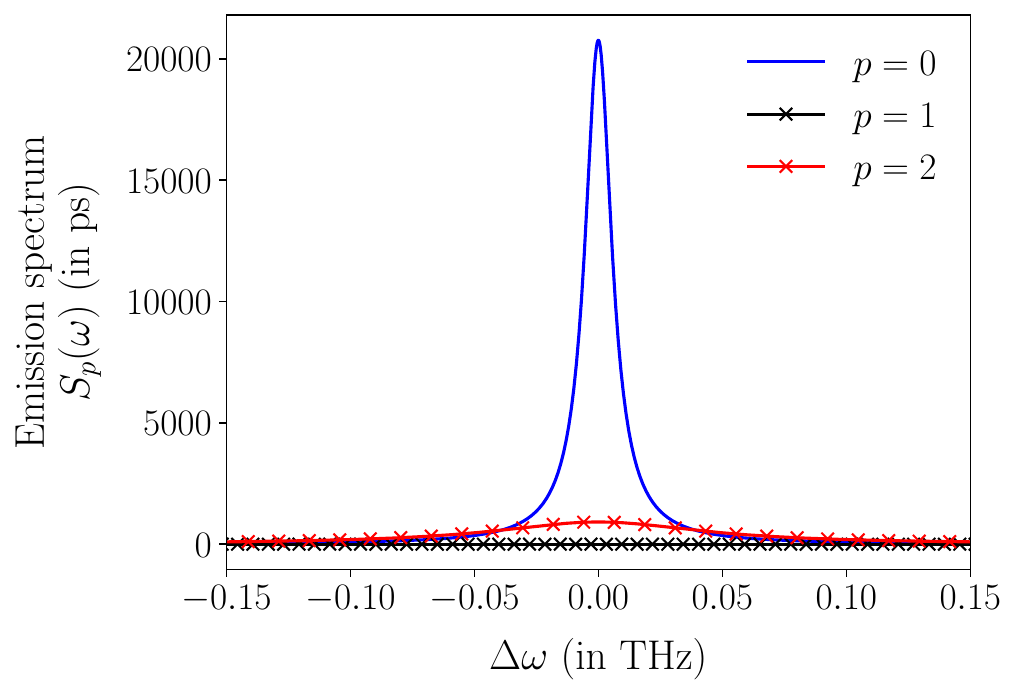} 
\caption{The emission spectrum of the first few photonic modes. 
The figure shows the spectral function $S_p(\omega)$, normalized with the initial steady state population $\langle\hat{a}^\dag_p(0)\hat{a}_p(0)\rangle$ = $\langle\hat{n}_p(0)\rangle$ at $t=0$), for the ground state mode ($p=0$) and the first and second modes ($p = 1, 2$). 
The pump is considered to be a Gaussian focused at the center, with a width equal to thrice the oscillator length $l_0$. The cavity cutoff frequency $\omega_0 \approx 520~\text{THz}$, with spacing between two adjacent cavity modes, $\Delta\omega = 1.7~\text{THz}$. The rate of loss of cavity photons is equal to $\kappa \approx 0.2~\text{THz}$ and the absorption ($\mathcal{A}_m$) and emission ($\mathcal{E}_m$) rates are calculated from experimental data~\cite{Nyman2017}. The loss of excitations outside the cavity modes is $\Gamma_\downarrow \approx 3 \times 10^{-5}~\text{THz}$ and pumping rate $\Gamma_\uparrow = 0.2\times\Gamma_\downarrow$. 
}
\label{fig_spectrum}
\end{figure}

\begin{figure}[h] 
\center
\includegraphics[width=0.42\textwidth]{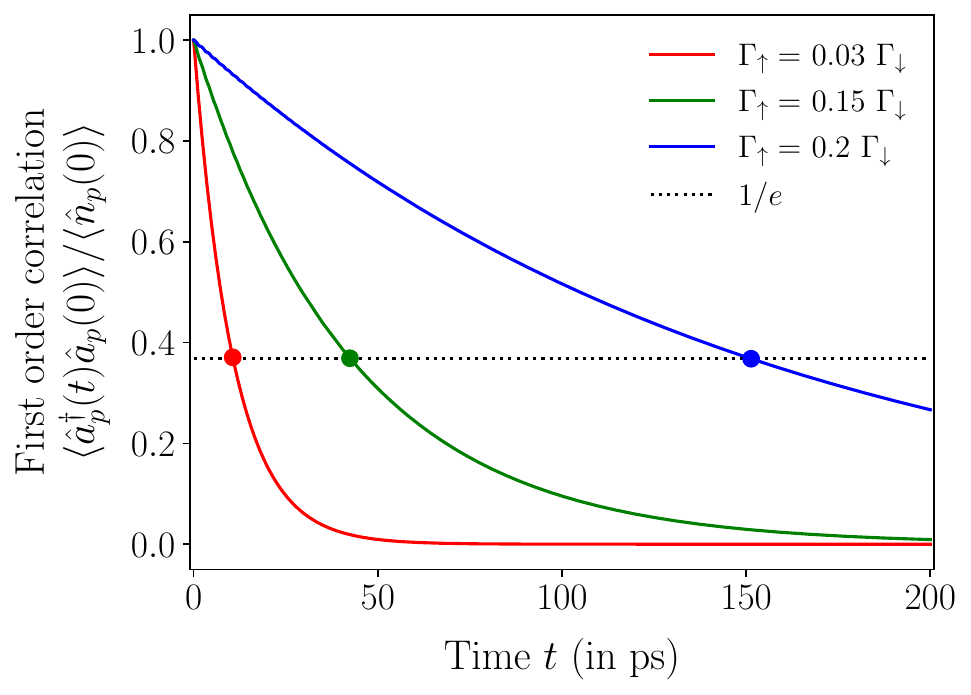} 
\caption{Time evolution of the first order correlation function. The figure shows the photon-photon correlation of the lowest energy state ($p = 0$) as a function of time $t$, normalized with the initial steady state population $\langle\hat{a}^\dag_p(0)\hat{a}_p(0)\rangle$ = $\langle\hat{n}_p(0)\rangle$ at $t=0$. The plots show the temporal behavior of the function for three different  
pump rates $\Gamma_\uparrow$ (in units of $\Gamma_\downarrow$). The bold circles mark the coherence time $\tau_0$ for the different pump rates.  All other system parameters are the same as in Fig.~\ref{fig_spectrum}.
}
\label{first_order}
\end{figure}

{The temporal coherence of the different photonic modes can be studied from the spectral function of the emitted light from the cavity~\cite{Walls1994}, given by the relation,
\begin{eqnarray}
 S_p(\omega) = \mathrm{Re}\bigg[\int_{-\infty}^{\infty} \langle\hat{a}_p^\dagger(t)\hat{a}_{p}(0)\rangle~ e^{i\omega t} dt \bigg],
\end{eqnarray}
which can be estimated by numerically solving for $\langle\hat{a}^\dag_p(t)\hat{a}_p(0)\rangle$ using
Eq.~(\ref{FULL22}).
Figure~\ref{fig_spectrum} shows the emission spectrum of the condensed ground state mode $p=0$, in comparison with the first and second excited modes, $p=1$ and $p=2$ respectively. The plots show that the ground state has a much narrower linewidth compared to the higher modes, which implies high temporal coherence in the condensed ground state mode.  
A notable point is the spectrum of the excited modes. The first excited mode has lower emission than the second excited mode, which implies that the second mode is more populated. 
This is due to the choice of a Gaussian pump spot focused at the center of the cavity, which tends to excite the even modes, with mode-functions that peak at the center. Moreover, the even modes also couple more strongly to the ground state~\cite{Tang2023}.}

Figure~\ref{first_order}, shows the time evolution of the photon-photon correlation of the ground state ($p=0$) for different pump powers. To facilitate the comparison, the correlation function is normalized with the initial steady state population at $t=0$. 
The photon-photon correlation decreases with time, but the rate of decrease reduces as the pump power is increased, which confirms the increased temporal coherence of the photons at higher pump powers.
Importantly, from the close to Lorentzian lineshape in Fig.~\ref{fig_spectrum}, it can be inferred that the decrease of the correlation function is exponential.
Thus the coherence time $\tau_p$ for cavity mode $p$ can be defined from the photon-photon correlation as
\begin{equation}
\langle\hat{a}^\dag_p(t)\hat{a}_p(0)\rangle = c_{p}(t) = c_{p}(0)\exp{[-{t}/{\tau_p}]}.
\end{equation}
The temporal coherence $\tau_0$ for the ground state for different pump powers are shown in Fig.~\ref{first_order} (bold circles), which marks the time at which the correlation function $c_{p}(t)$ drops to $1/e$ times its initial value $c_p(0)$.

\begin{figure}[t] 
\center
\includegraphics[width=0.42\textwidth]{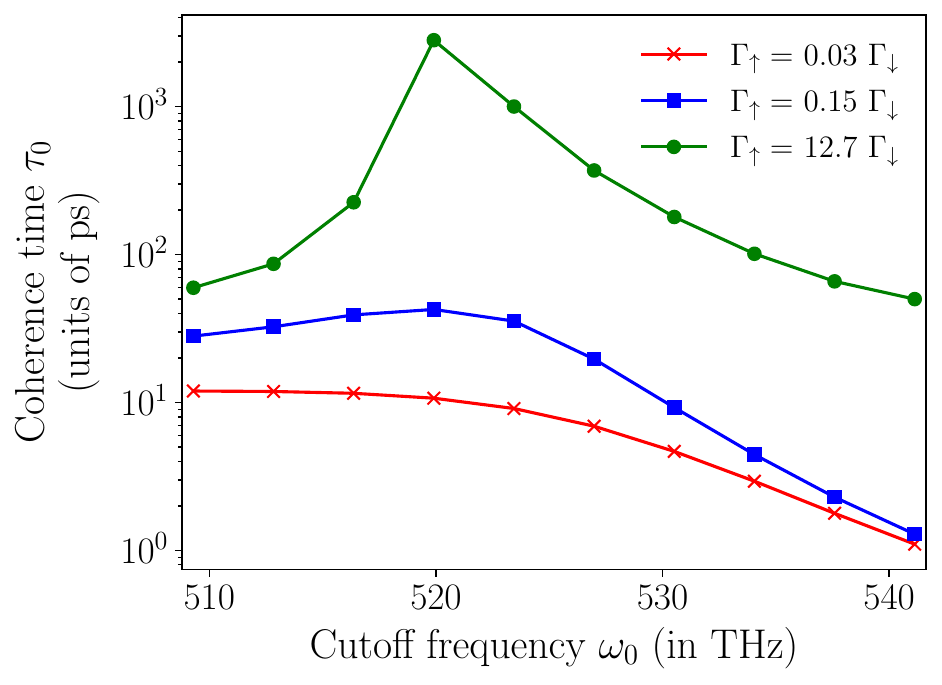} 
\caption{Variation of temporal coherence with cavity cutoff frequency. The plots show the coherence time of the lowest energy or ground state mode for different cutoff frequencies of the cavity, and for three different pumping rates $\Gamma_\uparrow$ (in units of $\Gamma_\downarrow)$. 
All other system parameters are the same as in Fig.~\ref{fig_spectrum}.
}
\label{tau_cutoff}
\end{figure}

Figure~\ref{tau_cutoff}, shows the variation of the coherence time $\tau_0$ of the lowest energy state with the cutoff frequency $\omega_0$ of the cavity. The cutoff frequency of the cavity is a critical system parameter that not only defines the energy of the lowest cavity mode, but also photon-energy dependent rates of absorption ($\mathcal{A}_p$) and emission ($\mathcal{E}_p$) of photons by the dye-molecules inside the cavity.  The figure shows that the coherence time is dependent on the cavity cutoff frequency, but does not vary monotonically with change in the cutoff frequency. This is either due to the nonmonotonic variation of thermalization inside the cavity across the chosen range of cutoff frequencies or strong mode competition for molecular excitations. The figure also shows how the coherence time of the ground state depends on the pump power, which controls the total number of photons in the cavity and therefore drives the photon condensation transition.\\

\section{Conclusion\label{conclusion}}

In this work, we derive an equation of motion for the first order correlation function or the photon-photon correlation for the photon gas inside a dye-filled microcavity. The nonequilibrium model takes into account a multimode photonic cavity, where finite inter-mode coherences are not completely ignored, which makes the calculations significantly more complex, but allows us to compute photon-photon correlations between different modes. Importantly, these relations allow the theoretical and computational investigation of temporal coherence of photon condensates that are consistent with actual experimental findings for a far wider set of parameters~\cite{Tang2023}.

The work opens the door to study more complex behavior of photon correlations and temporal coherence, especially in regimes where strong mode competition for excitation exists. A particular phenomenon of interest is to study how temporal coherence behaves in the regime of multimode condensate, where molecular excitations are clamped by a condensing mode and different modes compete to unclamp the excitation, leading to the phenomenon of decondesation~\cite{Hesten2018}. Moreover, other directions include a more comprehensive study of spatio-temporal correlations in photon condensation under the influence of non-stationary pump, where phenomena such as vortex-like structure formation\cite{Dhar2021}, and partially coherent light can be engineered.

\begin{acknowledgments}
The authors acknowledge financial support from the European Commission via the PhoQuS project (H2020-FETFLAG-2018-03) number 820392 and the EPSRC (UK) through the grants EP/S000755/1.
HSD acknowledges financial support from SERB-DST, India via a
Core Research Grant CRG/2021/008918 and the Industrial Research \& Consultancy Centre, IIT Bombay via grant (RD/0521-IRCCSH0-001) number 2021289.
\end{acknowledgments}

\appendix

\begin{widetext}
\section{Detailed calculation of coefficients\label{app1}}

The derivation of the coefficients in Sec.~\ref{coefficients}, which arise due to the different terms in the master equation given by Eq.~(\ref{master}), can be investigated more carefully. The first nontrivial term  is $L[\sigma_i^{\pm}]$, which represents the dynamics due to pumping and decay of molecules as governed by the rates $\Gamma_{\uparrow}^i$ and $\Gamma_\downarrow$, respectively. These are given by

\begin{align}
\sum_{\underline{n},\underline{s}}\dot{P}^{\underline{k}_p}_{\underline{n},\underline{s}}|\underline{n}-\underline{k}_p\rangle\langle\underline{n}| \otimes |\underline{s}\rangle\langle\underline{s}|&\overset{L[\sigma^+_i]}{\longrightarrow} -\frac{1}{2}\sum_{i,\underline{n},\underline{s}}\Gamma_{\uparrow}^i L[\hat{\sigma}^{+}_{i}] P^{\underline{k}_p}_{\underline{n},\underline{s}}|\underline{n}-\underline{k}_p\rangle\langle\underline{n}|\otimes |\underline{s}\rangle\langle\underline{s}|,\nonumber\\
&= -\sum_{\underline{n},\underline{s}}P^{\underline{k}_p}_{\underline{n},\underline{s}}|\underline{n}-\underline{k}_p\rangle\langle\underline{n}|\otimes \sum_{i,\forall s_i=0}\Gamma_{\uparrow}^i \bigg(|\underline{s}\rangle\langle\underline{s}|-
|\underline{s}+\underline{s}_i\rangle\langle\underline{s}+\underline{s}_i|\bigg).
\label{app:PA1}
\end{align}
%
%
%
The first term on the right hand side (RHS) of Eq.~(\ref{app:PA1}) contains the contribution from the $\{\sigma_i^-\sigma_i^+,\rho\}$ term of the operator $L[\sigma^+_i]$ which acts on $|\underline{s}\rangle\langle\underline{s}|$, if and only if $s_i = 0$ i.e., $\sigma_i^-\sigma_i^+|\underline{s}\rangle\langle\underline{s}| = |\underline{s}\rangle\langle\underline{s}|\sigma_i^-\sigma_i^+=\delta_{s_i,0}|\underline{s}\rangle\langle\underline{s}|$. 
Comparing the basis states $|\underline{n}-\underline{k}_p\rangle\langle\underline{n}|$ and $|\underline{s}\rangle\langle\underline{s}|$ on both sides, the coefficient from the first term is
$-P^{\underline{k}_p}_{\underline{n},\underline{s}}~\tilde{\Gamma}_\uparrow$, where the term $\tilde{\Gamma}_\uparrow=\sum_{i,\forall s_i=0} \Gamma^i_\uparrow$, which is the total pump rate at all unexcited sites in the state $|\underline{s}\rangle$.

\begin{figure}[t]
\centering
\label{S}
\includegraphics[width=0.6\textwidth]{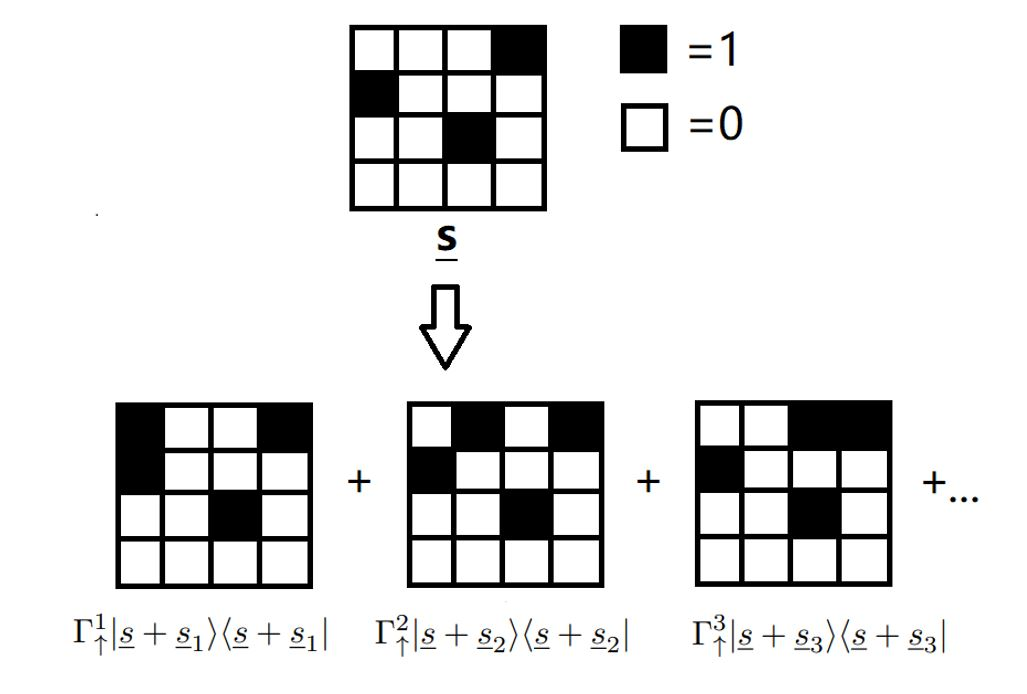} 
\caption{Visualization of the term $\sum_{i,\forall s_i=0} \Gamma_\uparrow^i$ as a square, where each block $\mathbf{r}_i$ denotes a position, and  black (white) implying the molecule at the position is excited (non-excited). For a specific $\underline{s}$ there are associated squares with one more excitation, give by a term $\Gamma_\uparrow^j$ which creates excitation at a new position $\mathbf{r}_j$, with new basis $|\underline{s}+\underline{s}_{j}\rangle\langle\underline{s}+\underline{s}_{j}|$.}
\label{grid1}
\end{figure}

The second term on the RHS arises due to the contribution from the $\sigma_i^+\rho\sigma_i^-$ term of $L[\sigma^+_i]$. Note that this term increases the excitation at locations where the molecules are unexcited or 
$\sigma_i^+|\underline{s}\rangle\langle\underline{s}|\sigma_i^- = \delta_{s_i,0}|\underline{s}+\underline{s}_i\rangle\langle\underline{s}+\underline{s}_i|$. This is illustrated in Fig.~\ref{grid1}. As such we obtain the term 
$\sum_{i,\forall s_i=0}\Gamma^i_\uparrow|\underline{s}+\underline{s}_{i}\rangle\langle\underline{s}+\underline{s}_{i}|$, which for any fixed $\underline{s}$ is a summation over all states with one more excitation than $\underline{s}$. 
Taking all $\underline{s}$ into account and noting that there will be no $|\underline{s}\rangle\langle\underline{s}|$, where $s_i=0~\forall i$, the total contribution of second term maybe rewritten by a simple change of indices 
\begin{eqnarray}
\sum_{\underline{n},\underline{s}}P^{\underline{k}_p}_{\underline{n},\underline{s}}|\underline{n}-\underline{k}_p\rangle\langle\underline{n}|\otimes \sum_{i,\forall s_i=0}\Gamma_{\uparrow}^i 
|\underline{s}+\underline{s}_i\rangle\langle\underline{s}+\underline{s}_i| \rightarrow 
\sum_{\underline{n},\underline{s'}}P^{\underline{k}_p}_{\underline{n},\underline{s}'-\underline{s}'_i}|\underline{n}-\underline{k}_p\rangle\langle\underline{n}|\otimes \sum_{i,\forall s'_i=1}\Gamma_{\uparrow}^i 
|\underline{s}'\rangle\langle\underline{s}'|.
\end{eqnarray}
The change of indices above is better illustrated in Fig.~\ref{grid2}.
\begin{figure}[h]
\centering
\includegraphics[width=0.6\textwidth]{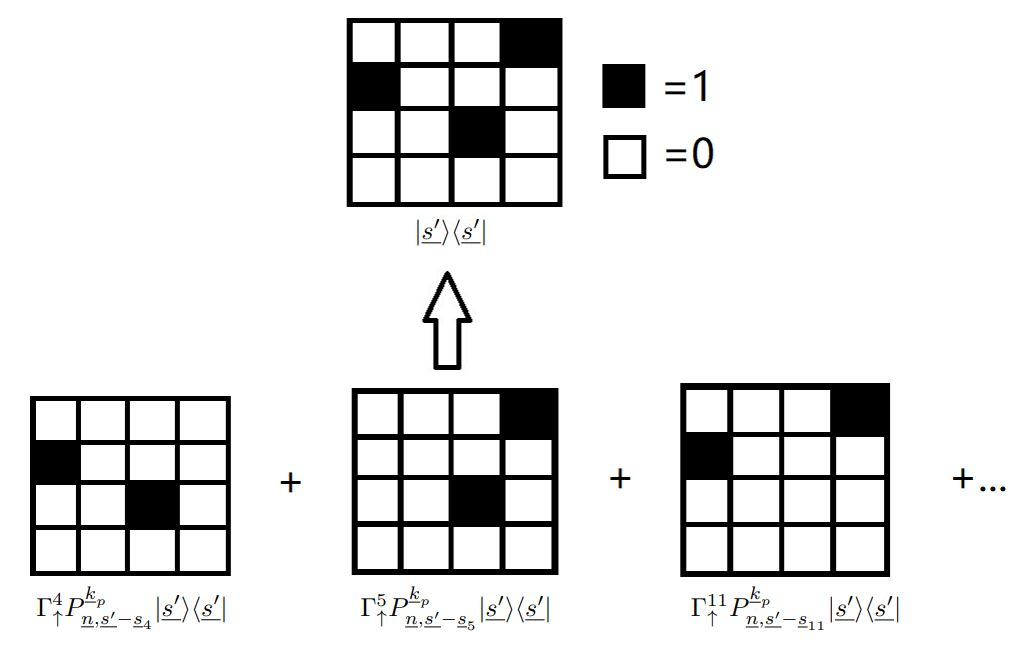} 
\caption{Visualization of  $|\underline{s}'\rangle\langle\underline{s}'|$ as a square, with different contributing terms coming from a set of squares, with one less excitation at position $\textbf{r}_j$ compared to $\underline{s}'$. Thus, each term contributes $\Gamma_\uparrow^j~P^{\underline{k}_p}_{\underline{n},\underline{s}'-\underline{s}_{j}}$.}
\label{grid2}
\end{figure}
%
%
%
%
%
Now by comparing terms in LHS and RHS with the same basis $|\underline{n}-\underline{k}_p\rangle\langle\underline{n}|\otimes |\underline{s}\rangle\langle\underline{s}|$, we obtain 
\begin{equation}
\dot{P}^{\underline{k}_p}_{\underline{n},\underline{s}}\overset{L[\sigma^+_i]}{=}\sum_{i,\forall s_{i}=1} \Gamma^i_\uparrow~P^{\underline{k}_p}_{\underline{n},\underline{s}-\underline{s}_{i}}-\tilde{\Gamma}_\uparrow~P^{\underline{k}_p}_{\underline{n},\underline{s}}.
\label{PA9}
\end{equation}
The contribution from decay to non-cavity modes governed by the $\Gamma_\downarrow$ term can be obtained in a similar manner.\\


Next, there are the contributions due to the absorption and emission processes, which are given by operators of the $\mathcal{A}_{m}[\hat{a}_m\hat{\sigma}^{+}_{i}\rho,\hat{a}^{\dagger}_{m'}\hat{\sigma}^{-}_{i}]$ and $\mathcal{E}_{m'}[\hat{a}^{\dagger}_{m'}\hat{\sigma}^{-}_{i}\rho,\hat{a}_m\hat{\sigma}^{+}_{i}]$, respectively. 
%
For the $\mathcal{A}_{m}$ term, the relevant relation between the coefficients is
\begin{equation}
\sum_{\underline{n},\underline{s}}\dot{P}^{\underline{k}_p}_{\underline{n},\underline{s}}|\underline{n}-\underline{k}_p\rangle\langle\underline{n}|\otimes |\underline{s}\rangle\langle\underline{s}|
~\overset{\mathcal{A}_{m}}{\longrightarrow}~\frac{1}{2}\sum_{m,m',i}\Psi_{m,m'}^i \mathcal{A}_{m'}
(\hat{a}_{m'}\hat{\sigma}^{+}_{i}\rho\hat{a}^{\dagger}_{m}\hat{\sigma}^{-}_{i}- \hat{a}^{\dagger}_{m}\hat{\sigma}^{-}_{i}\hat{a}_{m'}\hat{\sigma}^{+}_{i}\rho).
\label{PA11}
\end{equation}
%
%
Calculating the first term in RHS gives us the contribution from the term $\hat{a}_{m^{'}}\hat{\sigma}^{+}_{p}\rho~\hat{a}^{\dagger}_{m}\hat{\sigma}^{-}_{p}$:
\begin{align}
\frac{1}{2}\sum_{m,m',i}&\Psi_{m,m'}^i \mathcal{A}_{m'} \hat{a}_{m'} \hat{\sigma}^{+}_{i}\rho~\hat{a}^{\dagger}_{m}\hat{\sigma}^{-}_{i}\nonumber\\
&=\frac{1}{2}\sum_{m, m',i}\bigg(\sum_{\underline{n},\underline{s}}\Psi_{m,m'}^i\mathcal{A}_{m'}~P^{\underline{k}_p}_{\underline{n},\underline{s}}\sqrt{(n_{m'}-\delta_{p,m'})n_{m}}~|\underline{n}-\underline{k}_p-\underline{k}_{m'}\rangle\langle\underline{n}-\underline{k}_{m}|\otimes\delta_{s_{i},0}|\underline{s}+\underline{s}_{i}\rangle\langle\underline{s}+\underline{s}_{i}|\bigg)\nonumber\\
&=\frac{1}{2}\sum_{m, m',\underline{n},\underline{s}}\bigg[\mathcal{A}_{m'}\sqrt{(n_{m'}-\delta_{p,m'})n_{m}}~|\underline{n}-\underline{k}_p-\underline{k}_{m'}\rangle\langle\underline{n}-\underline{k}_{m}|\otimes\sum_{i,\forall s_i =1}\Psi_{m,m'}^i~P^{\underline{k}_p}_{\underline{n},\underline{s}-\underline{s}_{i}}|\underline{s}\rangle\langle\underline{s}|\bigg]\nonumber\\
&=\frac{1}{2}\sum_{m, m',\underline{n}',\underline{s}}\bigg[\mathcal{A}_{m'}\sqrt{(n'_{m'}-\delta_{p,m'}+\delta_{m,m'})(n'_{m}+1)}|\underline{n}-\underline{k}_p-\underline{k}_{m'}+\underline{k}_{m}\rangle\langle\underline{n}'|\otimes
\sum_{i,\forall s_i =1}\Psi_{m,m'}^i P^{\underline{k}_p}_{\underline{n}+\underline{k}_m,\underline{s}-\underline{s}_{i}}|\underline{s}\rangle\langle\underline{s}|\bigg],
\label{app:Am1}
\end{align}
where $\underline{n}-\underline{k}_{m}=\underline{n}'$ and the indices in the molecular states have been rearranged. \\



\noindent Similarly, the second term in RHS is the contribution due to  $\hat{a}^{\dagger}_{m}\hat{\sigma}^{-}_{i}\hat{a}_{m^{'}}\hat{\sigma}^{+}_{i}\rho$:
%
\begin{align}
\frac{1}{2}\sum_{p}&\Psi_{m,m'}^i \mathcal{A}_{m'} \hat{a}^{\dagger}_{m}\hat{\sigma}^{-}_{i}\hat{a}_{m'}\hat{\sigma}^{+}_{i}\rho\nonumber\\
&=\frac{1}{2}\sum_{m,m',\underline{n},\underline{s}} \bigg[\mathcal{A}_{m'}P^{\underline{k}_p}_{\underline{n},\underline{s}}\sqrt{(n_{m}+1-\delta_{m,p}-\delta_{m,m'})(n_{m'}-k_{m'})}|\underline{n}-\underline{k}_p-\underline{k}_{m'}+\underline{k}_{m} \rangle\langle \underline{n}|\otimes \sum_{i,\forall s_i=0}\Psi_{m,m'}^i |\underline{s}\rangle\langle\underline{s}| \bigg]
\label{app:Am2}
\end{align}

\noindent Combining the relations from Eqs.~(\ref{app:Am1}) and (\ref{app:Am2}), align the corresponding basis, the following expression for the coefficient is obtained
%
%
\begin{align}
\dot{P}^{\underline{k}_p}_{\underline{n},\underline{s}}\overset{\mathcal{A}_{m}}{\longrightarrow}
\frac{1}{2}\sum_{m, m'}&\bigg[\mathcal{A}_{m'}\sqrt{(n_{m'}-\delta_{p,m'}+1)(n_{m}+1)}\bigg(\sum_{i,\forall s_{i}=1} \Psi_{m,m'}^i P^{\underline{k}_p-\underline{k}_{m'}+\underline{k}_{m}}_{\underline{n}+\underline{k}_{m},\underline{s}-\underline{s}_{i}}\bigg)\nonumber\\
&-\mathcal{A}_{m'}P^{\underline{k}_p-\underline{k}_{m'}+\underline{k}_{m}}_{\underline{n},\underline{s}}\sqrt{(n_{m}+1-\delta_{m,p}-\delta_{m,m'})(n_{m'}-k_{m'})}\bigg(\sum_{i,\forall s_{i}=0}\Psi_{m,m'}^i\bigg)\bigg].
\label{PA16}
\end{align}
%
The contribution from the Hermitian conjugate and molecular emission, given by the rate $\mathcal{E}_{m'}$ in the master equation, can be calculated along similar lines.

\section{Derivation of equation of motion\label{app2}}

The detailed derivation of the equation of motion for the photon-photon correlation, using the coefficients arising from the different terms in the master equation, is presented. Again, the terms in the equation arising from $\hat{H}_{0}$ and $L[\hat{a}]$ are quite straightforward,

\begin{align}
\frac{d}{dt}\langle\hat{a}_p^\dagger(t)\hat{a}_{q}(0)\rangle &\overset{\hat{H}_{0}}{\longrightarrow}\sum_{\underline{n},\underline{s}} i\delta_{p}\sqrt{n_{p}}\dot{P}^{\underline{k}_p}_{\underline{n},\underline{s}}=i\delta_{p}\langle\hat{a}_p^\dagger(t)\hat{a}_{q}(0)\rangle \label{H0} \\
&\overset{L[\hat{a}]}{\longrightarrow}\sum_{\underline{n},\underline{s},m}\kappa\sqrt{n_p}[\sqrt{( n_{m}+1)(n_{m}+1-\delta_{p,m})} P^{\underline{k}_{p}}_{\underline{n}+\underline{k}_{m},\underline{s}}-(n_{m}-\frac{1}{2}\delta_{m,p})P^{\underline{k}_{p}}_{\underline{n},\underline{s}}]\\
&= -\frac{1}{2}\sum_{\underline{n},\underline{s}}\kappa\sqrt{ n_{p}}P^{\underline{k}_{p}}_{\underline{n},\underline{s}}{=}-\frac{\kappa}{2}\langle\hat{a}_p^\dagger(t)\hat{a}_{q}(0)\rangle,
\end{align}

as the $m\neq l$ terms cancels out in the calculation. Next, the contributions from the terms $L[\sigma^\pm_i]$ are considered:

%
\begin{equation}
\frac{d}{dt}\langle\hat{a}_p^\dagger(t)\hat{a}_q(0)\rangle \overset{L[\sigma_i^+]}{\longrightarrow} \sum_{\underline{n}}\sqrt{n_{p}}\sum_{\underline{s}}\bigg\{\bigg(\sum_{i,\forall s_{i}=1}\Gamma_\uparrow^i P^{\underline{k}_{p}}_{\underline{n},\underline{s}-\underline{s}_{i}}\bigg)-\sum_{i, \forall s_{i}=0}\Gamma_\uparrow^i P^{\underline{k}_{p}}_{\underline{n},\underline{s}}\bigg\}.
\label{eq98}
\end{equation}
Note in Eq.~(\ref{eq98}), $\sum_{\underline{s}}(\sum_{i, \forall s_{i}=1}\Gamma_\uparrow^i P^{\underline{k}_{p}}_{\underline{n},\underline{s}-\underline{s}_{i}})$ = $\sum_{\underline{s}'}c_{\underline{s}'}P^{\underline{k}_{p}}_{\underline{n},\underline{s}'}$,
is just a linear combination of $P^{k_{p}}_{\underline{n},\underline{s}'}$. 
For each $\underline{s}'$ on the right there exists only one term from left with $\underline{s}=\underline{s}'+\underline{s}_{i}~(\forall s_{i}=0)$, which then gives $c_{\underline{s}'} = \sum_{i,\forall s_{i}=0}\Gamma_\uparrow^i$. Including the expression for $c_{\underline{s}'}$ above, the following relation is obtained:  
\begin{equation}
\frac{d}{dt}\langle\hat{a}_i^\dagger(t)\hat{a}_{j}(0)\rangle\overset{L[\sigma_i^\pm]}{\longrightarrow} 0.
\label{A}
\end{equation}
A similar derivation can be done for contribution from the term $L[\sigma^-_i]$.

%

\noindent The contribution for the absorption and emission, given by the rates $\mathcal{A}_{m}$ and $\mathcal{E}_{m'}$, respectively, are considered next. First, the term $\mathcal{A}_{m}$ for the case $m=m'$:
%
\begin{align}
\frac{d}{dt}\langle\hat{a}_p^\dagger(t)\hat{a}_{q}(0)\rangle \overset{\mathcal{A}_{m}}{\longrightarrow} &\sum_{\underline{n},\underline{s}}\mathcal{A}_{p}\bigg[\sqrt{n_{p}+1}~n_{p}\bigg(\sum_{i,\forall s_{i}=1}|\psi_{p}^i|^{2} P^{\underline{k}_{p}}_{\underline{n}+\underline{k}_{m},\underline{s}-\underline{s}_{i}}\bigg)-(n_{m}-\frac{\delta_{m,p}}{2}) P^{\underline{k}_{p}}_{\underline{n},\underline{s}} \bigg(\sum_{i,\forall  s_{i}=0}|\psi_{p}^i|^{2}\bigg)\bigg]\nonumber\\
&+ \sum_{\underline{n},\underline{s},m\neq p}\mathcal{A}_{m}\bigg[(n_{m}+1)\sqrt{n_{p}}\bigg(\sum_{i,\forall  s_{i}=1}|\psi_{m}^i|^{2} P^{\underline{k}_{p}}_{\underline{n}+\underline{k}_{m},\underline{s}-\underline{s}_{i}}\bigg)-n_{m}\sqrt{n_{p}}P^{\underline{k}_{p}}_{\underline{n},\underline{s}} \bigg(\sum_{i,\forall s_{i}=0}|\psi_{m}^i|^{2}\bigg)\bigg],
\label{Am}
\end{align}
%
where $|\psi_{m}^i|^{2}$ is squared wavefunction of cavity mode $m$. Now, the last two lines in Eq.~(\ref{Am}) are all $m\neq p$ contributions, and by taking $\underline{n}'=\underline{n}+\underline{k}_{m}$, the last two lines become
%
\begin{align}
\sum_{\underline{n}',\underline{s},m\neq p} \mathcal{A}_{m} n_{m}'\sqrt{n_{p}}\bigg(\sum_{i,\forall s_{i}=1} |\psi_{m}^i|^{2} P^{\underline{k}_{m}}_{\underline{n}',\underline{s}-\underline{s}_{i}}\bigg)-\sum_{\underline{n},\underline{s},m\neq p} \mathcal{A}_{m} n_{m}\sqrt{n_{p}} P^{\underline{k}_{m}}_{\underline{n},\underline{s}} \bigg(\sum_{i,\forall s_{i}=0}|\psi_{m}^i|^{2}\bigg).
\label{Am2}
\end{align}
Similar to the calculations for $L[\sigma_i^\pm]$, 
the first bracket $\sum_{i,\forall s_{i}=0}|\psi_{m}^i|^{2}$ is transformed to $\sum_{i,\forall s_{i}=1}|\psi_{m}^i|^{2}$, and therefore cancel when $m\neq l$. Therefore only the $m=p$ terms remain
\begin{align}
\frac{d}{dt} \langle\hat{a}_p^\dagger(t)\hat{a}_q(0)\rangle\overset{\mathcal{A}_{m}}{\longrightarrow} -\frac{1}{2} \sum_{\underline{n},\underline{s}} \mathcal{A}_{p}\sqrt{n_{p}}~P^{\underline{k}_{p}}_{\underline{n},\underline{s}} \times\bigg(\sum_{i,\forall s_{i}=0}|\psi_{p}^i|^{2}\bigg).
\label{Am4}
\end{align}
%

\noindent For $m \neq m'$:

\begin{align}
\frac{d}{dt}\langle\hat{a}_p^\dagger(t)\hat{a}_q(0)\rangle &
\overset{\mathcal{A}_{m'}}{\longrightarrow} \sum_{\underline{n},\underline{s},m\neq m'} \frac{\mathcal{A}_{m'}}{2}\bigg\{\sqrt{n_{p}(n_{m'}-\delta_{p,m'}+1)(n_{m}+1)}\bigg(\sum_{i,\forall s_{i}=1}~\Psi_{m,m'}^i~P^{\underline{k}_i-\underline{k}_{m'}+\underline{k}_{m}}_{\underline{n}+\underline{k}_{m},\underline{s}-\underline{s}_i}\bigg)\nonumber\\
&-\sqrt{n_p(n_{m'}-\delta_{p,m'}+1-\delta_{m,m'})(n_{m}-\delta_{m,p})}\bigg(\sum_{i,\forall s_{i}=0}\Psi_{m,m'}^i\bigg)P^{\underline{k}_i-\underline{k}_{m'}+\underline{k}_{m}}_{\underline{n},\underline{s}}\nonumber\\
&+\sqrt{n_p(n_{m}-\delta_{m,l}+1)(n_{m'}+1)}\bigg(\sum_{i,\forall s_{i}=1} \Psi_{m,m'}^iP^{\underline{k}_i+\underline{k}_{m'}-\underline{k}_{m}}_{\underline{n}+\underline{k}_{m'},\underline{s}-\underline{s}_i}\bigg)
\nonumber\\ &
-\sqrt{n_pn_{m}(n_{m'}+1-\delta_{m,m'})}\bigg(\sum_{i,\forall s_{i}=0}\Psi_{m,m'}^i\bigg)P^{\underline{k}_i+\underline{k}_{m'}-\underline{k}_{m}}_{\underline{n}+\underline{k}_{m'}-\underline{k}_{m},\underline{s}}\bigg\}.\nonumber\\
&\overset{\mathcal{A}_{m'}}{\longrightarrow} \sum_{\underline{n},\underline{s},m\neq m'}\frac{\mathcal{A}_{m'}}{2}\bigg(\sum_{i,\forall s_{i}=0}\Psi_{m,m'}^i\bigg)
\bigg\{\sqrt{n_p(n_{m'}-\delta_{m',p}+1)(n_{m}+1)}
P^{\underline{k}_{p}-\underline{k}_{m'} +\underline{k}_{m}}_{\underline{n} +\underline{k}_{m},\underline{s}} \nonumber\\
&-\sqrt{n_p(n_{m'}-\delta_{m',p}+1-\delta_{m,m'})(n_{m}-\delta_{mi})}P^{\underline{k}_{p}-\underline{k}_{m'}+\underline{k}_{m}}_{\underline{n},\underline{s}}
+\sqrt{n_p(n_{m}-\delta_{mi}+1)(n_{m'}+1)}P^{\underline{k}_{p}+\underline{k}_{m'}-\underline{k}_{m}}_{\underline{n}+\underline{k}_{m'},\underline{s}}
\nonumber\\
&-\sqrt{n_pn_{m}(n_{m'}+1-\delta_{m,m'})}P^{\underline{k}_{p}+\underline{k}_{m'}-\underline{k}_{m}}_{\underline{n}+\underline{k}_{m'}-\underline{k}_{m},\underline{s}}\bigg\},
\label{Am6}
\end{align}

\noindent Rearrangement of some of the indices in the RHS of Eq.~(\ref{Am6}) gives us
\begin{align}
\frac{d}{dt}\langle\hat{a}_p^\dagger(t)\hat{a}_q(0)\rangle
&\overset{\mathcal{A}_{m'}}{\longrightarrow} \sum_{\underline{n},\underline{s},m\neq m'}\frac{\mathcal{A}_{m'}}{2}\bigg(\sum_{i,\forall s_{i}=0}\Psi_{m,m'}^i\bigg)
\bigg\{\sqrt{(n_{p}-\delta_{m,p})(n_{m'}-\delta_{p,m'}+1)n_{m}}
P^{\underline{k}_{p}-\underline{k}_{m'} +\underline{k}_{m}}_{\underline{n},\underline{s}}\nonumber\\
&-\sqrt{n_{p}(n_{m'}-\delta_{p,m'}+1-\delta_{m,m'})(n_{m}-\delta_{m,p})}P^{\underline{k}_{p}-\underline{k}_{m'}+\underline{k}_{m}}_{\underline{n},\underline{s}}
\nonumber\\
&+\sqrt{(n_{p}-\delta_{p,m'})(n_{m}-\delta_{m,p}+1)n_{m'}} P^{\underline{k}_{p}+\underline{k}_{m'}-\underline{k}_{m}}_{\underline{n},\underline{s}}\nonumber\\
&-\sqrt{(n_{p}-\delta_{p,m'}+\delta_{m,p})n_{m'}(n_{m}+1-\delta_{m,m'})}P^{\underline{k}_{p}+\underline{k}_{m'}-\underline{k}_{m}}_{\underline{n},\underline{s}}\bigg\}.
\label{Am7}
\end{align}


\noindent For $m\neq m'$, there can be two cases, either $m = p$ or $m \neq p$ for any $m'$. For the latter case, $\frac{d}{dt}\langle\hat{a}_p^\dagger(t)\hat{a}_{q}(0)\rangle = 0$. For $m = p$, the expression is
\begin{align}
\frac{d}{dt}\langle\hat{a}_p^\dagger(t)\hat{a}_{q}(0)\rangle &\overset{\mathcal{A}_{m'}}{\longrightarrow}
\sum_{\underline{n},\underline{s},m=p,m'}\frac{\mathcal{A}_{m'}}{2}\bigg(\sum_{i,\forall s_{i}=0}\Psi_{m',p}^i\bigg)\times\bigg\{n_{p}\sqrt{n_{m'}}P^{\underline{k}_{m'}}_{\underline{n},\underline{s}}-(n_{p}+1)\sqrt{n_{m'}}P^{\underline{k}_{m'}}_{\underline{n},\underline{s}}\bigg\},\nonumber\\
&=\sum_{\underline{n},\underline{s},m=p,m'}\frac{-\mathcal{A}_{m'}}{2}\bigg(\sum_{i,\forall s_{i}=0}\Psi_{m',p}^i\bigg)\sqrt{n_{m'}}P^{\underline{k}_{m'}}_{\underline{n},\underline{s}}.
\label{Am9}
\end{align}
Again, similar calculations exist for terms arising from emission, $\mathcal{E}_{m}$. 


\section{Semiclassical approximation.\label{app3}}

The semiclassical approximation used in obtaining the equation of motion in Sec.~\ref{sec:eom} is discussed in more detail. The approximation can be written as 
$P^{\underline{k}_{m}}_{\underline{n},\underline{s}}=P^{\underline{k}_{m}}_{\underline{n}}P_{\underline{s}}$,
where the probabilities for the photon and molecules are factorized, which implies that the two subsystems are uncorrelated. Moreover, 
$\sum_{\underline{s}}P_{\underline{s}}=1$, where $P_{\underline{s}}$ is probability of having excitation profile $\underline{s}$.
Under the semiclassical approximation we have
\begin{align}
\langle\hat{a}_p^\dagger(t)\hat{a}_{q}(0)\rangle=\bigg(\sum_{\underline{n}}\sqrt{n_{p}}~P^{\underline{k}_{p}}_{\underline{n}}\bigg)\times\bigg(\sum_{\underline{s}}P_{\underline{s}}\bigg)
=\sum_{\underline{n}}\sqrt{n_{p}}~P^{\underline{k}_{p}}_{\underline{n}}.
\label{MF1}
\end{align}

\noindent Applying the approximation in Eq.~(\ref{MF1}) to the equation of motion:
\begin{align}
\frac{d}{dt}\langle\hat{a}_p^\dagger(t)\hat{a}_{q}(0)\rangle &= (i\delta_{p}-\frac{\kappa}{2})\langle\hat{a}_p^\dagger(t)\hat{a}_{q}(0)\rangle -\sum_{m}\frac{\mathcal{A}_{m}}{2} \bigg(\sum_{\underline{n}}\sqrt{n_{m}}~P^{\underline{k}_{m}}_{\underline{n}}\bigg)\bigg(\sum_{\underline{s}}\sum_{i,\forall s_{i}=0} P_{\underline{s}}~\Psi_{m,p}^i\bigg)\nonumber \\
& +\sum_{m}\frac{\mathcal{E}_{m}}{2} \bigg(\sum_{\underline{n}} \sqrt{n_{m}}~P^{\underline{k}_{m}}_{\underline{n}}\bigg) \bigg(\sum_{\underline{s}}\sum_{i,\forall s_{i}=1} P_{\underline{s}}~\Psi_{m,p}^i\bigg)\nonumber \\
& = (i\delta_{p}-\frac{\kappa}{2})\langle\hat{a}_p^\dagger(t)\hat{a}_{q}(0)\rangle-\sum_{m}\frac{\mathcal{A}_{m}}{2}\langle\hat{a}_{m}^\dagger(t)\hat{a}_p(0)\rangle\bigg[\sum_{\underline{s}}\bigg(P_{\underline{s}}\sum_{i,\forall s_{i}=0}\Psi_{m,p}^i\bigg)\bigg]\nonumber \\
&+\sum_{m}\frac{\mathcal{E}_{m}}{2}\langle\hat{a}_{m}^\dagger(t)\hat{a}_p(0)\rangle \bigg[\sum_{\underline{s}}\bigg(P_{\underline{s}}\sum_{i,\forall s_{i}=1}\Psi_{m,p}^i\bigg)\bigg].
\label{MF2}
\end{align}
The term 
$\sum_{\underline{s}}(P_{\underline{s}}\sum_{i,\forall s_{i}=1}\Psi_{m,p}^i)$ can be transformed into $\sum_{i}\Psi_{m,p}^i (\sum_{\underline{s'},\forall \underline{s}_i=1}P_{\underline{s}'})$, where the last summation is over all $\underline{s}'$ with unity at position $i$. The sum $\sum_{s'}P_{\underline{s}'}$ is simply the total probability of net excitation of the $i^\text{th}$ molecule (i.e. the molecules at position $\mathbf{r}_i$). 
Using a similar argument, $\sum_{\underline{s}}(P_{\underline{s}}\sum_{i,\forall s_{i}=0}\Psi_{p,m}^i)$ can be changed to
$\sum_{i}\Psi_{p,m}^i (\sum_{s, \forall s_i = 0}P_{\underline{s}})$, where $\sum_{s}P_{\underline{s}}$ is the net probability of $i^{th}$ molecule being unexcited. Let $\underline{f}$ be a vector, where $f_{i}$ is the excitation fraction at position $\mathbf{r}_i$, which results in $\sum_{i}\Psi_{p,m}^i (\sum_{\underline{s'}}P_{\underline{s}'})=\sum_{i}f_{i}\Psi_{p,m}^i$ and  
$\sum_{i}\Psi_{p,m}^i (\sum_{s}P_{\underline{s}})=\sum_{i}(1-f_{i})\Psi_{p,m}^i.$ \\

\noindent The equation of motion of the photon-photon correlation under the semiclassical approximation is then given by, 
\begin{align}
\frac{d}{dt}\langle\hat{a}_p^\dagger(t)\hat{a}_{q}(0)\rangle =
(i\delta_{p}-\frac{\kappa}{2})\langle\hat{a}_p^\dagger(t)\hat{a}_{q}(0)\rangle-
\frac{1}{2}\sum_{m,i}\bigg({\mathcal{A}_{m}}\langle\hat{a}_{m}^\dagger(t)\hat{a}_p(0)\rangle (1-f_{i})\Psi_{p,m}^i -{\mathcal{E}_{m}}\langle\hat{a}_{m}^\dagger(t)\hat{a}_p(0)\rangle f_{i}\Psi_{p,m}^i\bigg).
\end{align}
\end{widetext}

\end{document}